# A Flexible Cryptographic Infrastructure for High-security SDR-based Systems

Peter Hillmann and Björn Stelte
Faculty of Computer Science
Universität der Bundeswehr München
85579 Neubiberg, Germany
Email: {peter.hillmann, björn.stelte}@unibw.de

*Abstract*—**Military software defined radio (SDR) systems are a major factor in future network-centric operations due to their flexibility and support for more capable radio communications systems. The inherent nature of software-based systems requires a more complex auxiliary infrastructure and multiple independent levels of security compared with typical systems: Secure booting of the SDR device, cryptographically signed software, real time operating platform software as well as radio applications. This technology raises new challenges with respect to the management. The largest impact on SDR deployments is due to the auxiliary cryptographic infrastructure for the security of the software life cycle and the cyclic update of the keys. Compared to conventional radio devices, the SDR system with the cryptographic infrastructure described in this paper reaches a higher security level and is more flexible. The advantage is the possibility to deploy trunked radio system and further waveforms, such as coalition wideband, which will be standardized in the future. Also it is possible to update cryptographic mechanisms. In this work, we analyze the requirements for a high secure SDR deployment and model the life cycle of the components of a deployed SDR node based on the Joint Program Executive Office (JPEO) Software Communication Architecture (SCA).**

I. INTRODUCTION

Software defined radio (SDR) devices for military usages are complex on behalf of the mandatory strong encryption functions, the configuration for ad-hoc networks and reconfiguration of multiple communication lines. To support operators by their hard work in front, a lot of devices have a simple structured management system in background with better working conditions. Tries is often to shift much functionality from the SDR device to the management to relieve the user. During an operation, the soldier has to do more important tasks than to deal with the configuration of the radio system. The interoperability with the management systems needs to be secured as well. Therefore, a key exchange is needed as well as other cryptographic sanctions and policies to realize military restrictions.

Strict demands on availability, reliability and security provide a further need for self-healing, self-configuration and practical usable concepts. With the life cycle in place, we devise an architecture for the auxiliary cryptographic infrastructures facilitating NATO standard operating procedure and assuring the user's security requirements. Different infrastructures need different update intervals and require special roles as well as security zones. Not every infrastructure needs its components and role-based management system. Several components of the auxiliary cryptographic infrastructure are allowed to last in the home country and have their own role-based management system. We present in this paper the entire path from allocation of the cryptographic material, over the generation of the keys, to the high secure distribution of operational and administrative data. The Operators are supported by adapted software components such as reminders for cyclic update intervals. The secure transfer of administrative data like new validated waveforms, real time operating system or other policies presented as well. There are a lot of challenges to the management, protocol design and network, which this paper identifies and analyses.

The paper is structured as follows. In Section II, an analysis of existing solutions and an overview of related work is presented. Section III describes the scenario for the scope of this paper with open problems. Section IV gives an analysis of the functional and design requirements based on the scenario. In Section V, the entire management structure is presented and all components as well as the interaction is described, followed by Section VI for the developed cryptographic container for data transportation. Section VII explains the cyclic update process, before the summary is given.

II. RELATED WORK

Over the past years, NATO has achieved interoperable secure voice communications through the use of the Narrow Band Secure Voice II (NBSV-II) series of equipments. These NBSV-II equipments operate over the homogeneous analog PSTN network using either common key material for NATO communications, or private key material for national communications. The NBSV-II devices are only capable of using pre-placed key material, with common key sets for all NATO entities requiring secure communications. Using separate key sets, the NBSV-II devices enable secure communications for national sovereign purposes. This type of operational scenario does not support modern day network enabled operations where dynamically negotiated session keys provide confidentiality, authentication and data integrity.

The Secure Communications Interoperability Protocols (SCIP) are a collection of application layer interoperability protocols designed to enable end-to-site and end-to-end secure voice and data communications across heterogeneous commercial and military bearer infrastructures and networks. Regardless of whether this communication is voice or data, an end-to-end data link is required. Secure voice communications use a data channel to transmit an AES encrypted digital representation of the speech signal across a concatenation of bearer networks. SCIP works over any media, such as RADIO, GSM, ATM, ISDN or SATCOM.

Both national as well as NATO secure voice requirements can be satisfied using SCIP-enabled products. The interoperability model created by the NBSV-II equipment is extended by the SCIP concept. The NBSV-II program was founded upon a homogeneous commercial infrastructure (PSTN) and established Communities of Interest (COIs) using interoperable equipment coupled with distinct key sets, NATO key for NATO communications and national key for national sovereign purposes. SCIP expands this concept by enabling secure end-to-end communications across a variety of bearers. Furthermore, any number of COIs can be established using multiple traffic encryption algorithms, both common and private protocols, and key materials. The critical point is that every nation must implement the identical call setup procedures, and at least one interoperable cryptographic mode.

NATO has historically achieved secure communications in the deployed segment using symmetric key material, also known as pre-placed key. Since NATO distributes symmetric key manually, this is a costly and labor intensive task. Asymmetric key negotiation provides each side of a secure communication with unique key material valid only for that session. In this type of key provisioning system, each side has a public and a private cryptographic key component. The strength of this system lies in the unique session keys as opposed to the common shared key as in a symmetric pre-placed key system. The Electronic Key Management System (EKMS) of NSA has been adopted as a standard for handling of key distribution and authentication. EKMS is required to update the master keys (both public and private) makes the asymmetric key negotiation much more flexible than the symmetric system, with dramatically lower life cycle costs. Periodic over the air re-keying (OTAR) is required by policy. There is currently no known method of extending this type of key negotiation beyond two parties, therefore, the only option for secure multiparty conferencing is to use preplaced key.

## III. SCENARIO

We consider a dynamic key management scenario where SDRs are used in a combat net radio operation or ad-hoc reconnaissance mission, and the pre-placed keys are not available to all participants prior to the missions. Each SDR has a minimum of two available channels, "x" and "y", where channel "x" is used for local net operations, (both voice and data) and channel "y" is used to relay communications traffic for other units. A secure communication can be established by temporarily "borrowing" bandwidth from channel "y" for a certificate based asymmetric key negotiation with a lead radio (or any radio in the network). Once this temporary secure link is established, the symmetric net key is transmitted to the new users. These new users can then join the secure communication as late entries using the SCIP protocols. Security is maximized in these scenarios because the strengths of both symmetric and asymmetric key material and authentication are employed. When the mission is over, then the session key expires and is destroyed, resulting in a less vulnerable overall system. SDR-based systems have the capability to support future communication techniques. The life cycle of an SDR device consists of three main phases: preparation, operation, recovery. During the preparation, the effort is to configure the device and load all necessary waveforms, policies and cryptographic data for nearly all possibilities during the operation. The problem consists of the different security level, which work in parallel. Auxiliary, some SDR devices have more than one unit for communication. Every unit can have different security levels. All SDR devices are able to encrypt the voice stream and data independently for every communication line before transmitting over the air. The radio system creates a secured overlay network. Beside the usage, the subject of the management for the time of the operation consists of monitoring and support. The recovery phase is to restore the SDR device in the sentence of a defect. New waveforms or future updates of existing software have to be distributed with attention to the life cycle. Before it can be done, all new data to the system have to be verified and encrypted for the transfer. Thus, a flexible adaption is needed. To develop an integrated management system all problems have to be addressed, which include the management and distribution of wideband waveforms and security policies as well as configurations files and new cryptographic algorithms. The main problem is often that such high security systems are not very practical.

## IV. REQUIREMENTS

The following part consists of a small security and usability analysis concentrating on the resilience of SDR deployments against various cyber-attack vectors. Furthermore the functional possibilities, such as the management for mobile SDR-based communication networks as multilingualism and strict guided operator interactions, are regarded. Flexibility, usability and accessibility have to be addressed with the same importance.

In the course of this threat analysis is of major importance for the mission fulfillment capability of the deployed SDR devices and of course for the end-user support helping to reduce the device's complexity for the operating user. The complexity for the end-user may not be higher than by legacy systems at a higher security level. Legacy systems are conventional systems, which are currently in use by the military and will work together with SDR systems. All components evince an intuitive usability and illustrating security functions. The effort for the required administrative infrastructure is reduced to a minimum in relation to the high security level. At the same time, attention is given on a high scalability cryptographic infrastructure referable to distribution of operational and administrative keys. The special requirement for military devices is the parallel processing of data with different classification levels. (Red-Red-Separation as well as NATO secret and national confidential)

In the following chapters, the components for a high-secure SDR System and the requirements on these components for possible NATO usage are presented.

## V. COMPONENTS

The management of the cryptographic infrastructures for SDR based systems has the problem of various update cycles for keys and certificates. Transferred information and data will be categorized to different security levels, which have special requirements. To face this challenge with corresponding and realizable concepts for a high security system, the architecture should be split in the administrative security management (ASM) and operational security management (OSM). Especially to reach the level NATO secret combined with a practical feasibility and usability. The ASM is responsible to provide the required data, for the configuration of the SDR device and for the planning of the time in usage alike afterwards. To transfer the large amount of information in a secure way and to be sure the correctness of the data a lot of cryptographic mechanisms are necessary. Updates for such data are not very often, but the data have mostly a high security categorization and need to be specially protected. The OSM is responsible for the small keys. The key and certificate material has to be transported and distributed on secure ways as well. To increase the security, the operational key for communication can build out of *n* independent keys from different OSM. In the following the individual components of the ASM and OSM are described. The ASM consists of radio security management system (RSMS) as highest trust instance and the radio network management system (RNMS) such as ones from the company Thales [1] or Elbit Systems [2], which is a compound of radio management systems (RMS) for the planning and configuration of the SDR device. The OSM includes the national generation and distribution management (NGDM), which generates the operational keys and the key distribution and management system (KDMS).

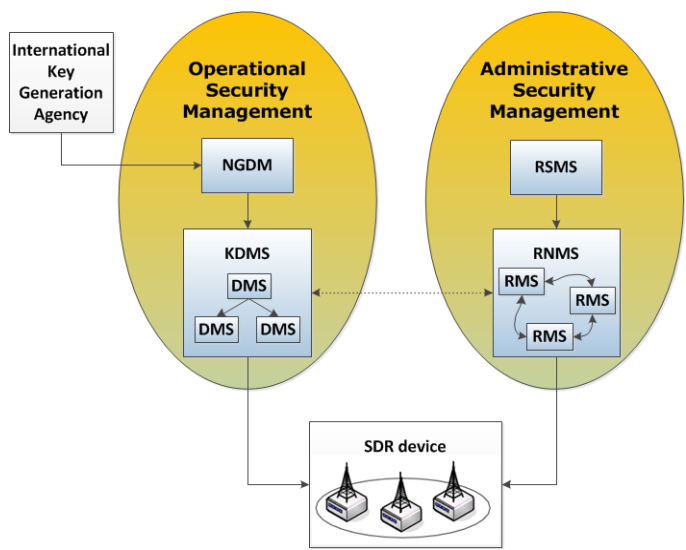

**Figure 1: Entire System**

### A. Radio Security Management System (RSMS)

This management system is the root of the security concept of the administrative management. As highest trust instance the system will generate certificates and keys for the distribution of administrative data. There exists a private cryptographic infrastructure between the RSMS and SDR device, where only these components have the possibility to share encrypted information. The transfer of the Container can be done by the military network or manually, for example USB-Sticks, because the information is strongly encrypted and only in hands of trusted persons for custody.

The RSMS is located in a restricted area without a network connection. It is designed as an offline available system. To reach a high security level, the SDR devices have to feature different elaborate security mechanisms such as authentication and authorization concepts with roles, passwords and capability lists to process high classified data. Access control lists are also possible, but have the disadvantage that for example new features require an update of the Access control lists in attention of security. Capability lists are easier to control for the role of a security officer and allow through white lists the access to security functions, which is more secure for such an important system. Nevertheless, modifications of rules must be flexible and quickly possible.

### B. Radio network management system (RNMS)

Confidential radio communication must be encrypted, which requires appropriate keys, certificates and other data. The problem is the availability at the SDR device. The RNMS has to distribute all configuration and planning data [3]. To construct a continuous chain of high level security, the SDR device and the multiple management systems have to follow rules. The RNMS is in charge to distribute the data from the

RSMS to the SDR devices by network or manually. The transferred containers are encrypted. The RNMS can identify the incoming data package from the RSMS on a header, but the RNMS is not able and do not need to decrypt the payload. Thereby the RNMS can underlie a lower security level than the RSMS.

The RNMS has its own lower private cryptographic infrastructure with SDR radio device for the distribution of further management information beside the packages of the RSMS. In addition to the standard tasks of management of the radio communication circles with heterogeneous devices and networks is "over the air management", observation of availability and proactive support by fault detection and support by waveform spreading under the geographical, situational and temporal conditions. A tracking feature for all radio systems in real time can be implemented as well. Future RNMS will support beside voice and data also video streams. The automatic mission planning and adaptive creation of configurations files will also generated necessary tasks for other components.

The planned authentication data have to be replicated locally to get a stable connection to the user management. But this replication has to be controlled by the RNMS itself. These management systems have also to administer data of different security classification levels. Sensitive information such as data packages from the RSMS combined with lower security level data as management information have to be processed by the RNMS in the same time.

The compound of radio management systems (RMS) is located directly in the military base at the operational area. The RMS synchronize their information and stored data over a peer-to-peer network with CORBA, XML or other practical formats so that all stations have the same base and view.

### C. National generation and distribution Management (NGDM)

For the part of the OSM the NGDM has the task to generate secret and cryptographic strong keys for the different algorithms to realize a secure connection between the allied forces. The amount of such keys can be very high, because, for example, the combatants have to change them sometimes daily. The electronic key generation system is located in the home country in a secure area. To increase the performance, multiple components can run in parallel. Generated keys are mostly used for operational aspects. The distribution of the keys can be done in two different ways. There exists the well-known Secure DTD2000 [4] from the NSA to transfer the keys personally by hand. The other way is the transfer over a secure internet connection, like a VPN Channel or SINA-Boxes [5, 6]. Normally, the keys will be transferred to a management station in the operational area. The data from the generation device have to be transferred manually with CD/DVD or unidirectional data diode [7] to a system with the encrypted connection to the military network.

During combined operations, the keys will be generated from a common organization such as NATO.

### D. Key distribution and management system (KDMS)

KDMS consists of multiple stations. The responsibility of the KDMS is limited to operational keys. The first station is located next to the NGDM and takes the keys from it to distribute them to the predefined children station until the data reaches the final node. This structure is chosen with the intention of scalability and flexibility for different amounts of cryptographic devices in operation.

The KDMS generate packages for bulk load methods, for example by Typhoons for the "Aircraft Crypto Variable Management Unit" [8]. Furthermore, the KDMS has to communicate with the RNMS and contrariwise for an adaptive planning of the encrypted communication circles. The KDMS need to know the available encryption algorithms and the necessary keys for the operation to assign the NGDM for generation.

### E. SDR device

To secure the SDR device against intrusion, it has tamper protection and is tempest proofed. Every communication channel at the SDR device has different memory spaces to reach multi independent levels of security. For the usage, the operator needs a dongle to authenticate with ownership plus password. Further, the dongle can be used to transfer data between SDR devices, KDMS and RNMS. The dongle has a direct secure connection to the SDR device, which is encrypted. The significance of such a dongle is limited to the administrative cryptographic data.

## VI. EXCHANGE CONTAINER

To transfer waveforms, keys, updates or other data objects, we developed a exchange container. The concept intends a cryptographic secure container for the distribution. Not every system has to get access to every file and the decryption has to be very performing, because of the hardware restrictions of the terminal.

The inner part of the container consists of the main content 1 to k data files to distribution and as the case may be the certificate to proof the signature. The storing of these worth to be protected information happened by a common file system. A solution for the separation is the distribution of the information on several containers. The decision depends beside the file size on the accessibility restriction of the receiver. The signature will be left unencrypted to reduce the data size against a cryptographic analysis and performance for checks. Due to the concept, the information inside the container is secured with the typical protection goals: confidentiality, integrity and authenticity.

It is not allowed to stack the container, even though it's possible. The reason for this restriction is the multiple usages of cryptographic algorithms in sequence. The constructed system is in the science well known as product cipher. Sequential Systems for concealment can't be proved at all, so that it has to be avoided by the management software [9].

The receiver of such a container first checks the signature with its pre stored or cached certificate. If the recipient isn't in possession, the stored certificate inside the inner container can be used. So the recipients don't have to be the certificate and it hasn't to be pre shared. If the check fails, the entire container has to be scrapped. For the decryption of the data object, the receiver needs to have the cryptographic data from the depending crypto infrastructure.

To reduce the calculation complexity during the encryption and decryption is a hybrid-system used. All data of the asymmetric encrypted transportation key will be added to the outer part of the container. In this way, for every SDR-based system different asymmetric keys are possible to choose. The two step keys prevent that the RNMS has to store for every SDR device a separate full sized Container with the data (for example platform software and waveform software). The Container will be generated by the RSMS only once and need to be stored at the RNMS also only once for all SDR devices in common. If a SDR device needs the data, the RSMS has to create the small sized Header with the key for the full sized package. The concept reduces the amount of transferred and stored data in the entire system. The advantage of the concept is that the amount of data which need to be pre-shared is very low.

## VII. KEY AND ALGORITHM UPDATE

The integrated algorithms and keys as well as key length and basic parameters are flexible and changeable to address the problem of cryptographic breaks and future developments as well as cryptographic analysis techniques. On an update, these information will be spread with a exchange container.

Based on the principle of Kerckhoffs, the security of a system should be based on the secrecy of the key and not on the secrecy of the encryption algorithms [10]. Experience shows that secret algorithms are proven to disclose themselves as weak or broken. By the principle of Kerckhoffs, it ensures that experts can analyze a method for possible weaknesses and highlight these. In addition, governmental institution such as the NIST in the USA provides general recommendations for the use of various encryption algorithms [11] for secured communication links, such as the use of AES to protect confidentiality, these algorithms are also public.

For our constructed cryptographic infrastructure, as authentication system will be used GMR [12], for the asymmetric encipherment CS [13] and symmetric encipherment Pseudo-one-time-pad [14] with s2-mod-n-Generator, but the algorithms are flex exchangeable. These algorithms are chosen on their high security level, cryptographic strong including active adaptive attacks.

The cyclic update intervals of the keys for the ASM are much longer than of the OSM. The cryptographic material of the ASM has a fixed period of validity, where in opposite the keys of the OSM have just an organizational validity. Furthermore, to have the possibility to react very fast on a cryptographic break or the suspicion, the SDR device should store for every infrastructure an additional key for standby.

## CONCLUSION

We have presented the concept of a flexible exchange container for SDR systems. The concept reduces the amount of transferred and stored data in the entire system. Due to its flexibility SDR related requirements are fulfilled.

In our work we summarize the critical attack vectors as well as their risks and impact on currently developed or deployed European SDR systems. The important part is the influence and additional expenditure on the key management system by a security leak. The outcome of this 'lessons learned' process enables us to provide alternative architectural and procedural elements to enhance military capability and resilience against cyber-attacks threatening the SDR deployment. The concluding of this work offers an outlook into future developments of military SDR systems and their auxiliary cryptographic infrastructures for NATO Secret communication systems. For future work, the entire system will be realized. Every component is programmed separately and after that the valid transfer of messages will be proofed.

## ACKNOWLEDGEMENT

The authors wish to thank the member of the Chair for Communication Systems and Internet Services at the Universität der Bundeswehr München, headed by Prof. Dr. Gabi Dreo Rodosek, for helpful discussions and valuable comments on previous versions of this paper.

This work was partly funded by Flamingo, a Network of Excellence project (ICT-318488) supported by the European Commision under its Seventh Framework Programm.

All authors are members of the Research Center Cyber Defense (CODE), which combines skills and activities of various institutes at the university, external organizations and IT security industry (for instance Cassidian or Giesecke and Devrient).